\documentclass{article}[12pt]
\usepackage{amsmath}
\newcommand{\set}[1]{\{ #1 \}}
\newcommand{\seq}[1]{\langle#1\rangle}

\newcommand{\nulls}{\mathit{null}}

\newcommand{\concat}{\mathop{\mathrm{concat}}}

\begin{document}
\title{Defining and composing big state machines}

\author{Victor Yodaiken}
\maketitle
\begin{abstract}
A sequence function alternative representation of state machines.
\end{abstract}
\section{Introduction}
State machines are usually presented in terms of a set of events $E$, a set of states $S$, and a map $\delta:S\times A\to S$. Alternatively we can define a state variable as a function  the set of finite sequences over $E$ so that for any sequence $s$:
\[ x = F(s)\]
The function $F$ is an alternative presentation of a state machine in a way made precise in section \ref{sec:machines}. We can define a number of state variables (maybe a large number) all depending on the same sequence of events to describe a complex system.   For example, given some sequence $s$ that defines the current state, we might require that the number of processes that are executing code in a critical region never be more than one - where $Processes$ and $ProgramCounter(p)$ and $Critical$ might all be state dependent.
 \[ |\set{p\in Process : ProgramCounter(p)\in Critical}| \leq 1\]
 
Section \ref{sec:def} shows techniques for defining and composing sequence functions. Section \ref{sec:machines} shows the correspondence between sequence functions and standard state machine presentations. Examples of applications are in other papers. This work comes from a long process of attempting to refine initial intuitions about the utility of recursive sequence presentations of state machines and the use of general state machine products to model composition and parallel/concurrent computation\cite{DBLP:journals/corr/abs-0907-4169} and \cite{DBLP:journals/corr/abs-0805-2749}

\section{Sequence functions\label{sec:def}}
Many useful sequence functions can be defined by primitive recursion on sequences.
Let $\nulls$ be the empty sequence and if $s$ is a sequence and $a\in E$, let $sa$ be the sequence obtained by appending $a$ to $s$. 
\begin{equation}G(\nulls)=x_0,\mbox{     } G(sa) = g(a, G(s))\end{equation} 
defines $G$ on every finite sequence - assuming $g$ is a previously defined function.

For example here is a mod $k$ counter (or we could leave off the mod $k$ and have an infinite state counter):
\[\begin{array}{l} C_k(\nulls)=0,\\ C_k(sa) =\begin{cases} C_k(s)+1\bmod k&\mbox{if }a=increment;\\
                                                                             0&\mbox{if }a=reset;\\
                                                                             C_k(s)&\mbox{otherwise}\end{cases}\end{array}\]
Standard function composition ``hides state":
\[D_k(s)  =\begin{cases}1&\mbox{if }C_k(s) \neq 0;\\
                                                                             0&\mbox{if otherwise.}\end{cases}\]
 A simple composition produces a tuple of parallel sequence functions:
\begin{equation} G(s) = (G_1(s),\dots G_n(s))\label{direct}\end{equation}
For example to count mod $10$ and mod $100$ and mod $1000000$ at the same time: 
\[F(s) = (C_{10}(s),\quad C_{100}(s), C_{100000}(s))\]
 so that $F(s)=(x,y,z)$ is a triple showing the three parallel counters.  

To make components communicate requires a second level of recursion that is analogous to "simultaneous recursion" in classic primitive recursive function theory\cite{Peter}. In this case, recursion used to produce sequences for each component from the ``global" sequence. For example,  connect 2 mod $k$ counters in a series so that counter $1$ counts units, and counter $2$ increments only when counter $1$ rolls over to 0. Take the sequence $s$ and define maps to $s_1$ and $s_2$ for the two components so that 
\[H(s) = (C_k(s_1), C_k(s_2))\]
Now define the relationship between $s$ and $s_1$ and $s_2$ be this:
\begin{itemize}
\item  when we append a $reset$ to $s$, append a $reset$ to both $s_1$ and $s_2$ ,
\item  when we append $increment$ to $s$, append $increment$ to $s_1$ and 
\begin{itemize}
\item  leave $s_2$ unchanged (if $C_k(s_1) < k-1$) or
\item  append $increment$ to $s_2$ if $C_k(s_1)= k-1$.
\end{itemize}\end{itemize}
 Then $H(s)=(n,m)$ indicates a count of $(n + m*k)\bmod k^2$.  

More generally, suppose that we have a collection of $n$ components each described by $G_i:E_i^*\to X_i, (0<i \leq n)$ and a set of global events $E$. A map $g$  has to be defined to specify the interaction between components so that  $g(i,a,x_1\dots x_n)= r_i$ gives the sequence of events component $i$ sees when a single event $a$  is appended to the global sequence and the component current state values are given by $x_1 \dots x_n$.
 The function $g$ determines a map $g^*(i,s)= s_i$ by recursion. Let $s \concat r$ be the sequence obtained by concatenating sequences $s$ and $r$.  We set $g^*(i,\nulls)=\nulls$ and then if $g(i,s)=s_i$ we define  $g^*(i,sa) = s_i\concat r_i$ where $r_i = g(i,a, G_1(s_1)\dots G_n(s_n))$.

\begin{equation}\begin{array}{l}
G(s)= (G_1(g^*(1,s))\dots G_n(g^*(n,s)))\\
\mbox{and }\begin{array}{l} g^*(i,\nulls)=\nulls\mbox{ and}\\
g^*(i,sa)= g^*(i,s)\concat g(a,G_1(g^*(1,s)),\dots G(g^*(n,s)))\end{array}\end{array}\label{eq:feedback}\end{equation}

\section{State machines \label{sec:machines}}
Obviously, finite state machines are an important class but I do not here assume state machines are finite.

 A Moore type state machine is $M=(X, E,S,\sigma_0,\delta,\lambda)$  where $\sigma_0\in S$ is the ``start state'' and $\delta:S\times E\to S$ is the transition function $\lambda: S\to X$ is the output map and $\lambda(\sigma)$ is the output of the state machine in state $\sigma$. A standard state machine can be considered to be a more machine where $\lambda(\sigma)=\sigma$. 

Given Moore machine $M= (X,E,S,s_0,\delta,\lambda)$ let $E^*$ be the set of finite sequences over $E$ including $\nulls$ and let \[\delta^*(\sigma,\nulls)=\sigma\]
 and \[\delta^*(\sigma,sa)= \delta(\delta^*(\sigma,s),a).\] Let  $M^*$ be defined by
\[M^*(s)=\delta^*(\sigma_0,s)\]
Then $\lambda_M(M^*(s))$ is the output of $M$ in the state reached by following $s$ from the initial state.

Say $M$ is an implementation of  $G:E^*\to X$ if and only if  $\lambda_M(M^*(s))=G(s)$ for all $s\in E^*$.

 Say $G$ is finite state if and only if there is a $M$ that implements $G$ where the state set of $M$ is finite.

If $E$ and $X$ are finite sets  and $g:E\times X\to X$ then $G$ defined by $G(\nulls)=x_0\in X$ and $G(sa)=g(a,G(s))$ is finite state.

  Suppose that $M_1,\dots M_n$ implement $G_1,\dots G_n$ and $M_i=(X_i,E_i,S_i,\sigma_{0,i},\delta_i,\lambda_i)$.   Define $G$ using definition form \ref{eq:feedback}. Then we can construct a product of the $M_i$ that implements $G$ as follows:
\begin{eqnarray*}
\mbox{The state set }S= S_1\times \dots S_n\\
\mbox{The output map }\lambda((\sigma_1\dots \sigma_n)=(\lambda_1(\sigma_1),\dots \lambda_n(\sigma_n))\\
\mbox{The transition map }\delta(\sigma,a) = (\dots \delta_i^*(\sigma_i,g(i,a, \lambda(s))\dots )\mbox{ where }\sigma=(\sigma_1,\dots \sigma_n)
\end{eqnarray*}
Clearly, $G$ is finite state if all the $G_i$ are finite state. The state machine product here is well known. See for example \cite{Gecseg}.  

There is a type of state machine product often called a ``cascade'' product where the flow of information does not include any cycles. Note that if for all $s\neq \nulls$ we have $\delta^*(\sigma,s)\neq \sigma$ then the state machine has only trivial cycles. This has some structural implications \cite{pin}. But cascades describe state machine products with only trivial cycles - something different, although there may well be a relationship. The Krohn-Rhodes theorem \cite{arbib} relates cascade products of state machines to simple groups and the well known Jordan-Hölder theorem. 

In a definition of type \ref{eq:feedback} consider whether $g(i,x_1\dots,x_n)$ depends on the $j^{th}$ element $x_j$ or not. For example, if we define $g(i,a,x)=\seq{a}$ then $g$ and $i$ do not depend on any of the $x_j$. If there is a partial order $R$ on $ \set{1\dots ,n}$ so that for each $g$ and $i$ do not depend on any $j$ with $i R j$, then say that the composite system is a ``cascade''.

\bibliography{all}

\begin{thebibliography}{Yod09}

\bibitem[Arb68]{arbib}
Michael~A. Arbib.
\newblock {\em Algebraic theory of machines, languages, and semi-groups}.
\newblock Academic Press, 1968.

\bibitem[Gec86]{Gecseg}
Ferenc Gecseg.
\newblock {\em Products of Automata}.
\newblock Monographs in Theoretical Computer Science. Springer Verlag, 1986.

\bibitem[Pet67]{Peter}
Rozsa Peter.
\newblock {\em Recursive functions}.
\newblock Academic Press, 1967.

\bibitem[Pin86]{pin}
J.E. Pin.
\newblock {\em Varieties of Formal Languages}.
\newblock Plenum Press, New York, 1986.

\bibitem[Yod08]{DBLP:journals/corr/abs-0805-2749}
Victor Yodaiken.
\newblock State and history in operating systems.
\newblock {\em CoRR}, abs/0805.2749, 2008.

\bibitem[Yod09]{DBLP:journals/corr/abs-0907-4169}
Victor Yodaiken.
\newblock Primitive recursion and state machines.
\newblock {\em CoRR}, abs/0907.4169, 2009.

\end{thebibliography}
\bibliographystyle{alpha}
\end{document}